%% file: main.tex
\documentclass[conference]{IEEEtran}
\IEEEoverridecommandlockouts
\usepackage{cite}
\usepackage{amsmath,amssymb,amsfonts}
\usepackage{amsthm}
\usepackage[lined,linesnumbered,boxed,vlined,ruled]{algorithm2e}
\usepackage{amsmath}
\usepackage{amssymb}
\usepackage{graphicx}
\usepackage{textcomp}
\usepackage[table]{xcolor}
\usepackage{setspace}
\usepackage{xspace}
\usepackage{booktabs}
\usepackage{url}
\usepackage[normalem]{ulem}
\usepackage{enumitem}
\usepackage{balance}
\usepackage{float}
\usepackage{subcaption}
\usepackage{caption}
\usepackage{multirow}
\usepackage{diagbox}
\usepackage{wasysym}

\definecolor{mygrey}{RGB}{230,230,240}

\def\BibTeX{{\rm B\kern-.05em{\sc i\kern-.025em b}\kern-.08em
    T\kern-.1667em\lower.7ex\hbox{E}\kern-.125emX}}

\newcommand\revise[1]{{\color{black}#1}}

\newcommand{\sysname}{\textsf{StreamShield}\xspace}
\newcommand{\godel}{G{\"o}del\xspace}

\newtheorem*{definition*}{Definition}

\SetKwRepeat{Do}{do}{while}
\SetKwComment{Comment}{// }{}

\begin{document}

\title{\sysname: A Production-Proven Resiliency Solution for Apache Flink at ByteDance}

\author{\IEEEauthorblockN{Yong Fang, Yuxing Han$\dagger$, Meng Wang, Yifan Zhang, Yue Ma, Chi Zhang}
\IEEEauthorblockA{\textit{Data Platform, ByteDance Inc.}\\
\thanks{$\dagger$ Dr. Yuxing Han is the corresponding author, hanyuxing@bytedance.com.}
}
}
\maketitle

\begin{abstract}

Distributed Stream Processing Systems (DSPSs) form the backbone of real-time processing and analytics at ByteDance, where Apache Flink powers one of the largest production clusters worldwide. Ensuring resiliency, the ability to withstand and rapidly recover from failures, together with operational stability, which provides consistent and predictable performance under normal conditions, is essential for meeting strict Service Level Objectives (SLOs).
However, achieving resiliency and stability in large-scale production environments remains challenging due to the cluster scale, business diversity, and significant operational overhead.
In this work, we present \sysname, a production-proven resiliency solution deployed in ByteDance’s Flink clusters. Designed along complementary perspectives of the engine and cluster, \sysname introduces key techniques to enhance resiliency, covering runtime optimization, fine-grained fault-tolerance, hybrid replication strategy, and high availability under external systems.
Furthermore, \sysname proposes a robust testing and deployment pipeline that ensures reliability and robustness in production releases.
Extensive evaluations on a production cluster demonstrate the efficiency and effectiveness of techniques proposed by \sysname. 
\end{abstract}


\input{sections/introduction}

\input{sections/preliminaries}

\input{sections/runtime}

\input{sections/fault_tolerance}

\input{sections/startup}

\input{sections/high-availability}

\input{sections/experiment}

\input{sections/conclusion}

\balance

\bibliographystyle{IEEEtran}
\bibliography{ref}

\end{document}

%% file: sections/introduction.tex
\section{Introduction}

At ByteDance, Distributed Stream Processing Systems (DSPSs) 
form the backbone of real-time data processing and analytics, supporting various business operations and user-facing applications.
By enabling the ingestion and processing of massive data streams collected from users, devices, and services, DSPS enables low-latency computation and the timely derivation of actionable insights.
Our production environment employs Apache Flink~\cite{carbone2015apache} as the core stream processing engine, operating one of the largest DSPS clusters worldwide.

In large-scale and business-critical streaming environments, failures are an unavoidable reality. 
In such contexts, resiliency~\cite{chandramouli2017shrink,hwang2005high,xing2006providing}, the ability to withstand and rapidly recover from adverse events such as {node failures, network partitions, and workload surges},
is paramount for meeting strict Service Level Objectives (SLOs)~\cite{mei2020turbine}.
Ensuring resiliency in production Flink clusters is essential for maintaining availability, controlling latency, and preserving correctness despite unpredictable runtime conditions. 
A complementary factor in achieving this goal is operational stability, which maintains consistent and predictable performance under normal operating conditions, thereby reducing performance fluctuations and limiting the need for recovery interventions.

However, preserving the resiliency of the Flink cluster at ByteDance presents significant challenges arising from the scale, diversity, and complexity of its operational environment:

\noindent \textbf{Cluster Scale: Managing High-Volume Stateful Streaming at Hyperscale.}
Our production-grade stream processing systems encounter {unprecedented} scalability challenges.
So far, the Flink cluster sustains over 70,000 concurrent streaming jobs and manages more than 11 million resource slots.
At this scale, hardware failures, network instabilities, software bugs, and resource contentions become routine rather than exceptional.
These disruptions 
often manifest as task failures, prolonged task lag, and repeated job restarts,
collectively degrading system-wide processing stability.
Ensuring the resiliency of Flink jobs under these frequent fault conditions, while simultaneously minimizing operational overhead and mitigating business impact, represents a significant challenge.

\begin{table*}[t]
    \centering
    \renewcommand{\arraystretch}{1.6}
    \scalebox{0.7}{
        \begin{tabular}{|>{\columncolor[HTML]{EFEFEF}}c|c|c|}
            \hline
            \diagbox[width=12em, height=5em, innerleftsep=1pt, innerrightsep=5pt]{\textbf{Recovery Latency}}{\textbf{Data Completeness}} & \textbf{Strict Data Completeness Required} & \textbf{Tolerant of Minor Data Loss} \\
            \hline
            \textbf{Sub-Second Recovery} &
            \begin{tabular}[c]{@{}c@{}}Ideally Preferred for All Business Scenarios\end{tabular} &
            \begin{tabular}[c]{@{}c@{}}Latency-critical Services \\ (e.g., Target Advertising \& Real-time Recommendation)\end{tabular} \\
            \hline
            \textbf{Sub-Minute Recovery} &
            \begin{tabular}[c]{@{}c@{}}Revenue-critical Services  (e.g., Monetization Metrics: TikTok tipping) \\ Data Synchronization Pipelines (e.g., streaming ETL workloads)
            \end{tabular} &
            \begin{tabular}[c]{@{}c@{}}Log-driven Analytical Pipelines \\(e.g., Short-video Engagement Statistics: Views, Clicks)\end{tabular} \\
            \hline
            \textbf{Hour-Level Recovery} &
            \begin{tabular}[c]{@{}c@{}} Data Warehousing Jobs (e.g., Scheduled Operational Dashboards)\end{tabular} &
            \begin{tabular}[c]{@{}c@{}}Not applicable; Prone to  System Malfunctions\end{tabular} \\
            \hline
        \end{tabular}
    }
    \caption{Heterogeneity across Business Scenarios.}
    \vspace{-2em}
    \label{tab:heterogeneity}
\end{table*}

\noindent \textbf{Business Diversity: Heterogeneous Workloads with Varying SLOs:}
The Flink cluster underpins various internal business applications with heterogeneous workloads.
Table~\ref{tab:heterogeneity} presents a taxonomy of business scenarios characterized by different requirements \textit{w.r.t} recovery latency and data completeness. 
Sub-second recovery with strict completeness represents the ideal target across all business scenarios, whereas some latency-critical services are willing to relax completeness guarantees in exchange for ultra-low latency. 
Revenue-critical workloads and data synchronization pipelines mandate complete data delivery with recovery guarantees within sub-minute bounds to safeguard business correctness. In contrast, log-driven analytical pipelines can tolerate minor losses at the same latency level. At the opposite end of the spectrum, hour-level recovery is sufficient for offline data warehousing jobs like scheduled operational dashboards when strict completeness is preserved, but becomes infeasible in loss-tolerant scenarios due to the risk of systemic malfunction. 
These heterogeneous workloads demand adaptive resiliency mechanisms that balance latency and correctness in diverse large-scale streaming environments.

\noindent \textbf{Operational Overhead: Managing Complex External Dependencies.}
Managing resiliency at the scale of ByteDance’s Flink deployment is complicated by its extensive reliance on external services and infrastructure components.
These dependencies span fundamental components such as HDFS~\cite{hadoop_hdfs_2025}, Apache Zookeeper~\cite{hunt2010zookeeper}, and Gödel~\cite{xiang2023godel}, and extend to numerous external systems that collectively form a multifaceted runtime environment.
Failures or performance anomalies in any of these external systems may propagate into Flink’s runtime, leading to frequent job restarts, task failovers, and on-call interventions.
Empirical evidence indicates that the cluster experiences thousands of such recovery events on a daily basis, significantly increasing operational complexity.
Ensuring stable 
performance under these conditions requires advanced fault isolation and recovery mechanisms that can effectively mitigate the impact of external dependencies.

\noindent \textbf{Our Approach.} 
To address the above challenges, we propose \sysname, a production-proven resiliency solution for Apache Flink deployed at ByteDance, designed along three \textit{complementary} perspectives: engine, cluster, and release.
From the engine perspective (Section~\ref{sec:runtime}), \sysname enhances \textit{runtime resiliency} through adaptive optimizations and \textit{fine-grained fault-tolerance} mechanisms to narrow the recovery scope.
In addition, it \textit{accelerates job startup} by optimizing job parsing and task deployment, and incorporates a HotUpdate mechanism that reuses existing resources to reduce restart overhead and improve deployment efficiency.

Among these techniques, some address general-purpose resiliency in hyperscale clusters with heterogeneous workloads.
Examples include \textit{Adaptive Shuffle}, which enables dynamic and load-aware data redistribution, and \textit{Autoscaling}, which adjusts resource allocation in response to workload surges.
Other techniques are tailored for specific business scenarios.
For example, for scenarios where limited data loss is tolerable, \textit{WeakHash} relaxes the strict key-to-task binding by mapping each key to a bounded set of candidate tasks and dynamically selecting the execution task, thereby diffusing hot keys across multiple tasks to alleviate 
data skew.
In parallel, \textit{Single-task Recovery} narrows the recovery scope from execution regions to individual tasks, allowing failed tasks to be restored without restarting their neighbors, which substantially reduces recovery latency and minimizes disruption to unaffected operators.

From the cluster perspective (Section~\ref{sec:high-availability}), \sysname adopts a hybrid replication strategy to balance recovery latency and operational overhead. It employs passive replication as the default mechanism and 
selectively
switches to active replication for latency-critical workloads with stringent availability requirements.
Beyond replication, \sysname strengthens robustness against failures in external dependencies 
through high-availability configuration, backoff and retry strategies, and multi-layer fault tolerance techniques.
These mechanisms localize failures from external environments, prevent cascading disruptions, and preserve cluster-wide stability under different infrastructure dynamics.

From the release perspective (Section~\ref{sec:release}), \sysname incorporates a robust testing and deployment pipeline. 
This pipeline incorporates systematic chaos testing~\cite{rosenthal2020chaos,shekhar2024chaos} and performance benchmarking to validate the stability of new versions under fault-prone and high-load conditions, ensuring reliability prior to production rollout.
By bridging controlled validation with real-world deployment, the pipeline not only validates resiliency under fault-prone and high-load conditions but also reduces operational overhead by automating defect detection and preventing costly post-deployment failures.

To summarize, this paper makes the following contributions through the design and deployment of \sysname:
\begin{itemize}
	\item \textbf{Engine-level Resiliency}: We propose techniques that accelerate job startup,  adapt execution dynamically via runtime optimizations, and enhance fault tolerance through fine-grained recovery mechanisms.
	\item \textbf{Cluster-level Resiliency}: We develop a hybrid replication strategy combining passive and active replication, with high-availability mechanisms for external dependencies. 
    \item \textbf{Release-level Resiliency}: We build a comprehensive testing and deployment pipeline that integrates chaos testing, micro- and macro-benchmarking, and online probe tasks.
	\item \textbf{Empirical Validation}: Extensive production-scale evaluation at ByteDance, demonstrating that \sysname significantly improves system resiliency, reduces operational overhead, and sustains stable performance under heterogeneous and failure-prone environments.

\end{itemize}

%% file: sections/preliminaries.tex
\section{Preliminaries}

This section outlines the architecture and basic concept of Apache Flink, followed by a formal definition of the resiliency problem in distributed stream processing.

\subsection{Apache Flink}

\begin{figure}[tb]
  \centering
  \includegraphics[scale=0.35]{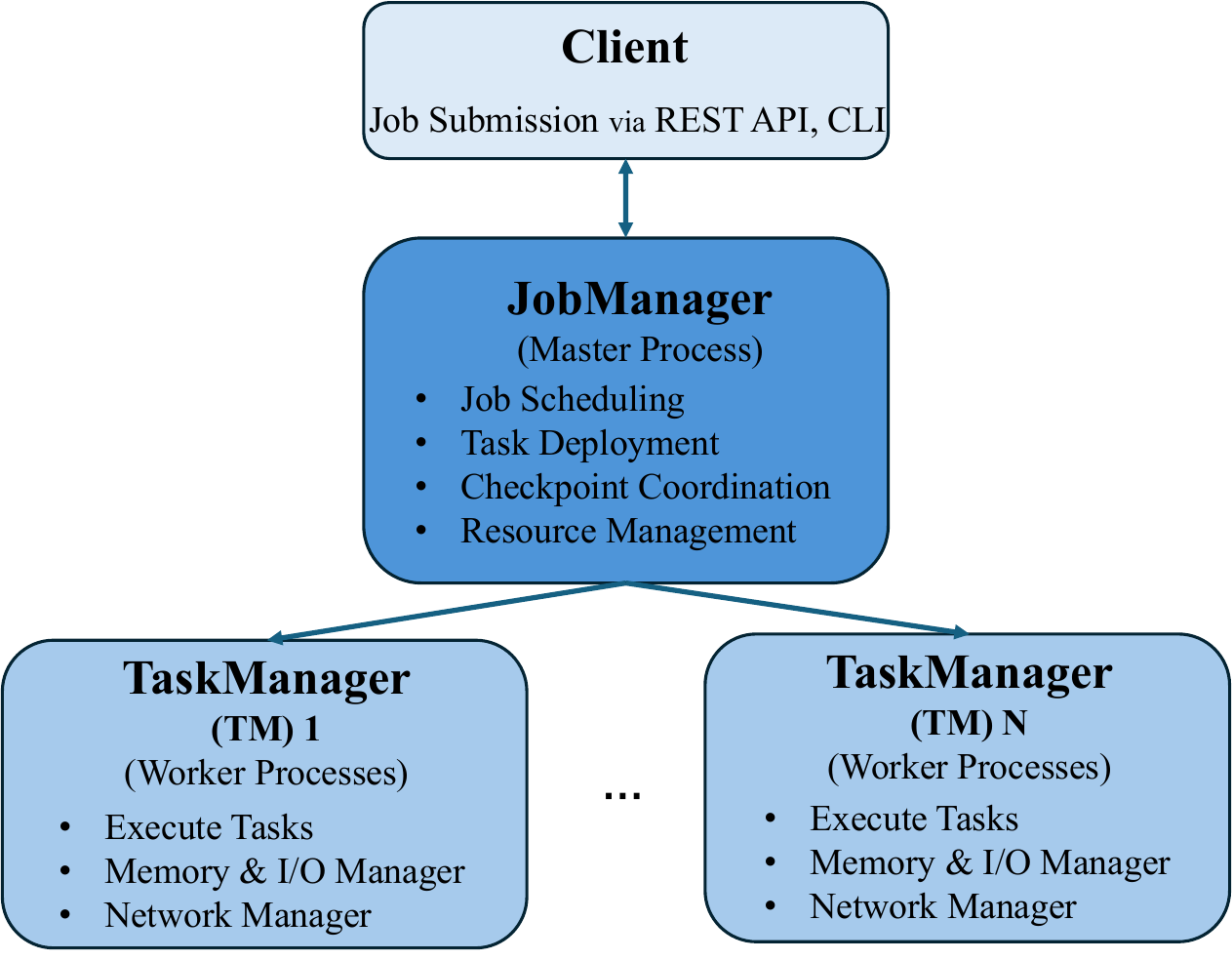}
  \caption{The Architecture of Apache Flink.}
  \vspace{-2em}
  \label{fig:flink_architecture}
\end{figure}

\noindent \uline{\textbf{Architecture.}}
Apache Flink is an open-source distributed stream processing framework for high-throughput, low-latency, and stateful computations over unbounded and bounded data streams.
As shown in Fig.~\ref{fig:flink_architecture}, its architecture is composed of three core components: the \textit{Client}, the \textit{JobManager}, and a set of \textit{TaskManagers}.
The Client compiles the job into a logical \textit{dataflow graph}~\cite{akidau2015dataflow,murray2013naiad,carbone2015lightweight}, and submits the job to JobManager,  which converts it into a physical execution plan.
The JobManager then identifies available resources by querying registered TaskManagers and assigns tasks to suitable task slots.
Once deployment instructions are issued, TaskManagers instantiate the corresponding tasks and begin processing data streams, while the JobManager concurrently coordinates checkpointing and monitors execution progress.

\noindent \uline{\textbf{Job, Operator and Task}} 
In Flink, a \textit{job} represents the complete streaming application submitted by the user, expressed as a logical dataflow of operators. An \textit{operator} denotes a high-level transformation or computation (e.g., map, filter, join) that defines the logical processing semantics. 
During execution, each operator is parallelized into multiple \textit{tasks}, which are the actual runtime units scheduled on TaskManagers. Tasks are assigned to task slots, consume input streams, maintain local state, and emit output records. 
Thus, a job consists of a directed acyclic graph of operators, and each operator is further materialized into one or more tasks, which together form the physical execution graph.

\noindent \uline{\textbf{Checkpoints v.s. Savepoints.}}
Apache Flink supports fault tolerance and state management through two key concepts: \textit{checkpoints} and \textit{savepoints}. \textit{Checkpoints} are automatically triggered, lightweight, and typically \textit{incremental} snapshots of the job state used for recovery in case of failures, ensuring exactly-once processing guarantees~\cite{flink_checkpoints}. 
In contrast, \textit{savepoints} are manually triggered and represent \textit{full} snapshots of the state that are externally stored and version-independent, primarily used for operational purposes such as job upgrades, migration, or long-term state preservation~\cite{flink_savepoints}.
Savepoints complement checkpoints by providing greater stability and flexibility, making them indispensable for managing the lifecycle of stateful Flink jobs.

\subsection{Problem Formalization}

Let a distributed stream processing job $J$ be represented as a directed acyclic graph 
$G=(V, E)$, where each vertex $v \in V$ is an operator instance (task) and each edge $e\in E$ denotes a data dependency between operators. The job executes over a set of physical nodes $N$, with each node hosting one or more tasks.
We define \textbf{resiliency} as the system’s ability to maintain SLO compliance in the presence of abnormal events 
$E=\{e_1, e_2,..., e_k\}$, where $e_i$ represents an adverse runtime condition such as hardware faults, network partitions, software crashes, or workload surges.

Given a service-level objective $\mathcal{S} = (\gamma, \lambda_{max}, \tau_{max})$, where $\gamma\in \{\textsf{full}, \textsf{partial}\}$ indicates whether the workload requires complete data processing or can tolerate limited data loss,  $\lambda_{max}$ is the maximum tolerated end-to-end latency, $\tau_{max}$  is the maximum allowable recovery time after an abnormal event.
Our problem can be stated as: 

\begin{definition*}[Resiliency Problem]
For any occurrence of $e_i \in E$ during the execution of $J$ over $N$, design mechanisms that ensure:
\begin{align*}
completeness\_requirement(J) = \gamma, \\
latency(J) \leq \lambda_{max},\\
recovery\_time(J, e_i) \leq \tau_{max}
\end{align*}

\noindent while minimizing additional resource overhead and operational cost.
\end{definition*}

%% file: sections/runtime.tex
\section{Resilient Runtime and Adaptive Execution} \label{sec:runtime}
From an engine perspective, achieving resiliency in large-scale stream processing requires not only 
adaptive runtime capabilities that sustain stable execution under dynamic workloads
but also 
robust fault-tolerance mechanisms.
To this end, \sysname introduces a set of mechanisms for runtime optimization, fine-grained fault tolerance. Besides, several techniques are proposed to speed up the job startup.

\subsection{Runtime Optimization}\label{sec:runtime_sub}
To ensure compliance with service-level objectives under three key constraints, \sysname incorporates several runtime optimization techniques that adapt to dynamic workloads, data skew, and fluctuating system conditions.
Specifically, \textit{Adaptive Shuffle} dynamically balances load across downstream tasks, \textit{WeakHash} alleviates data skew in key-based partitioning, and \textit{Autoscaling} adjusts operator parallelism in response to workload variability.

\begin{figure}[t]
\centering
\subfloat[Rebalance.]{
\label{fig:rebalance}
\raisebox{15pt}{\includegraphics[trim=0.4cm 0.4cm 0cm 0.4cm,scale=0.4]{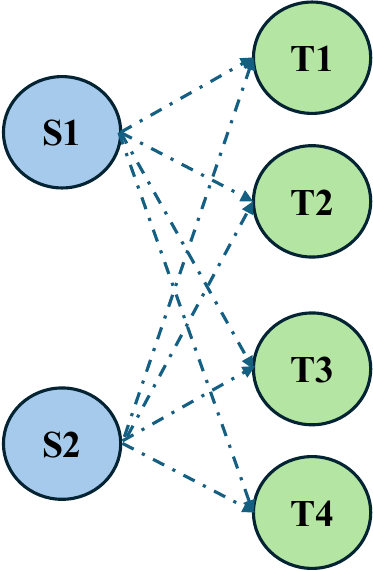}}
}
\subfloat[Rescale.]{
\label{fig:rescale}
\raisebox{15pt}{\includegraphics[trim=0.4cm 0.4cm 0cm 0.4cm, scale=0.4]{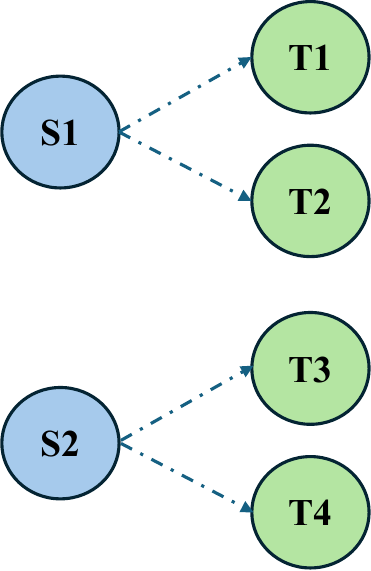}}
}
\subfloat[Group-Rescale.]{
\label{fig:group-rescale}
\includegraphics[trim=0.4cm 0.4cm 0cm 0.4cm, scale=0.26]
{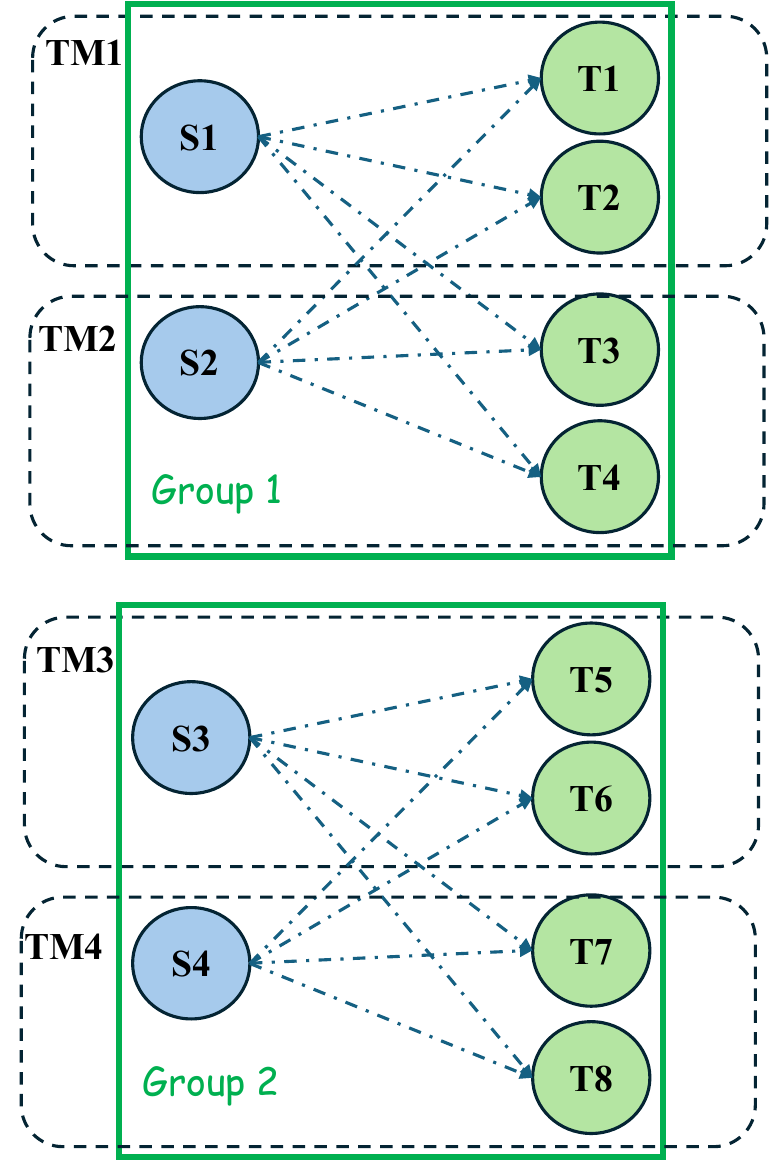}
}
\caption{{Shuffle Strategies Optimization.}}
\vspace{-2em}
\end{figure}

\noindent \uline{\textbf{Backlog-based Shuffle.}} \label{sec: adaptive_shuffle}
Flink provides two adaptive shuffle strategies for redistributing data among parallel tasks: \textit{Rebalance} and \textit{Rescale}.
The \textit{Rebalance} strategy (Fig~\ref{fig:rebalance}) redistributes records uniformly across all downstream subtasks in a round-robin manner, achieving even load distribution independent of operator parallelism configurations. 
By comparison, the \textit{Rescale} strategy (Fig.~\ref{fig:rescale}) restricts data exchange to a \textit{localized} subset of downstream subtasks aligned with the upstream parallelism structure; each upstream subtask communicates with a fixed range of downstream subtasks.

Since both Rebalance and Rescale distribute records using a round-robin algorithm, they implicitly assume homogeneous downstream processing abilities. 
However, this assumption frequently breaks down in real-world deployments due to node heterogeneity and workload imbalance~\cite{rivetti2016online}. 
To address this limitation, we propose \textit{backlog-based shuffle}, a dynamic and downstream-load-aware data redistribution strategy that builds upon Flink’s credit-based flow control mechanism~\cite{kruber2019deepdive}, wherein each downstream task issues a bounded number of buffer \textit{credits} to its upstream counterpart to regulate transmission.
In backlog-based shuffle, a configurable credit threshold is maintained per channel (i.e., the physical communication path between an upstream and a downstream task); once the threshold indicates overload, data is diverted away from the congested channel.
By proactively steering records toward underutilized nodes, this adaptive routing policy mitigates backpressure propagation, strengthens checkpoint reliability, and improves overall system stability and resource utilization.

\noindent \uline{\textbf{Group-Rescale.}} 
To enhance the flexibility of upstream-to-downstream redistribution under the Rescale strategy, we propose an optimization termed \textit{Group-Rescale}.
Unlike standard Rescale, where each upstream task sends data to a fixed subset of downstream tasks based solely on parallelism alignment, Group-Rescale adopts a group-based communication paradigm.
Specifically, the upstream and downstream tasks are partitioned into multiple disjoint groups, each comprising a subset of upstream and downstream tasks. 
Within each group, upstream tasks are fully connected to all downstream tasks, and data shuffling is strictly confined to the group boundary.

\revise{This strategy effectively addresses scenarios where co-located tasks on a single TaskManager collectively become stragglers. 
As illustrated in Fig.~\ref{fig:group-rescale}, when tasks S1, T1, and T2 are deployed on the same slow TaskManager, the backlog-based shuffle mechanism becomes ineffective since all co-located tasks suffer from similar backpressure. 
With Group-Rescale, data redistribution can occur across TaskManagers within the same group (e.g., TM1 and TM2 in Group 1), thereby expanding the communication scope beyond a single node. This enables healthy upstream tasks to bypass straggling downstream tasks, achieving more balanced load distribution and preventing localized bottlenecks from propagating through the entire dataflow.

Conceptually, Group-Rescale strikes a balance between Rebalance and Rescale. While Rebalance provides complete flexibility in data redistribution at the cost of increased network and scheduling overhead, and Rescale minimizes communication overhead but suffers from limited adaptability, Group-Rescale achieves a middle ground. It preserves the lightweight coordination of Rescale while partially relaxing its communication constraint, enabling controlled cross-TaskManager exchanges within groups.}

\noindent \uline{\textbf{WeakHash.}}
Traditional key-based partitioning in Flink enforces strong consistency by routing all records with the same key to a single downstream task.
While this design is necessary for stateful streaming operators requiring deterministic aggregation, it often leads to severe load imbalance when a small number of hot keys dominate the input distribution.
Such skew results in processing lag, backlog accumulation, and resource underutilization, which ultimately degrade the end-to-end performance of large-scale streaming jobs.

To mitigate this situation, \sysname introduces \textit{WeakHash}, a partitioning strategy that deliberately relaxes the strict key-to-task binding under the observation that certain scenarios can tolerate minor data loss.
A defining characteristic of WeakHash is its dynamic selection of execution targets for each record.
Rather than deterministically mapping each key to a single TaskManager, WeakHash associates the key with a \textit{bounded} group of candidate TaskManagers,  from which the actual downstream task is selected using lightweight dispatch strategies, such as load-aware selection guided by runtime metrics.
This design diffuses the load of heavily skewed keys across multiple tasks, thereby preventing hotspots and improving overall workload balance.

{Beyond scenarios where data loss is permissible,} WeakHash proves particularly effective in data warehousing workloads involving dimension table lookups, where fact streams are augmented with auxiliary attributes retrieved from distributed caches or external key-value stores (e.g., ABase~\cite{abase2025sigmod}).
These workloads are highly susceptible to key skew, as a small set of popular dimension keys (e.g., frequently accessed user or product IDs) may dominate the query distribution. 
Note, dimension lookups are typically idempotent and do not mutate the external table. Therefore, the relaxation of per-key consistency does not compromise correctness. Instead, it diffuses storage access pressure across multiple connections and stabilizes query latency under heavy skew.

\begin{figure*}[tb]
  \centering
  \includegraphics[width=0.95\linewidth]{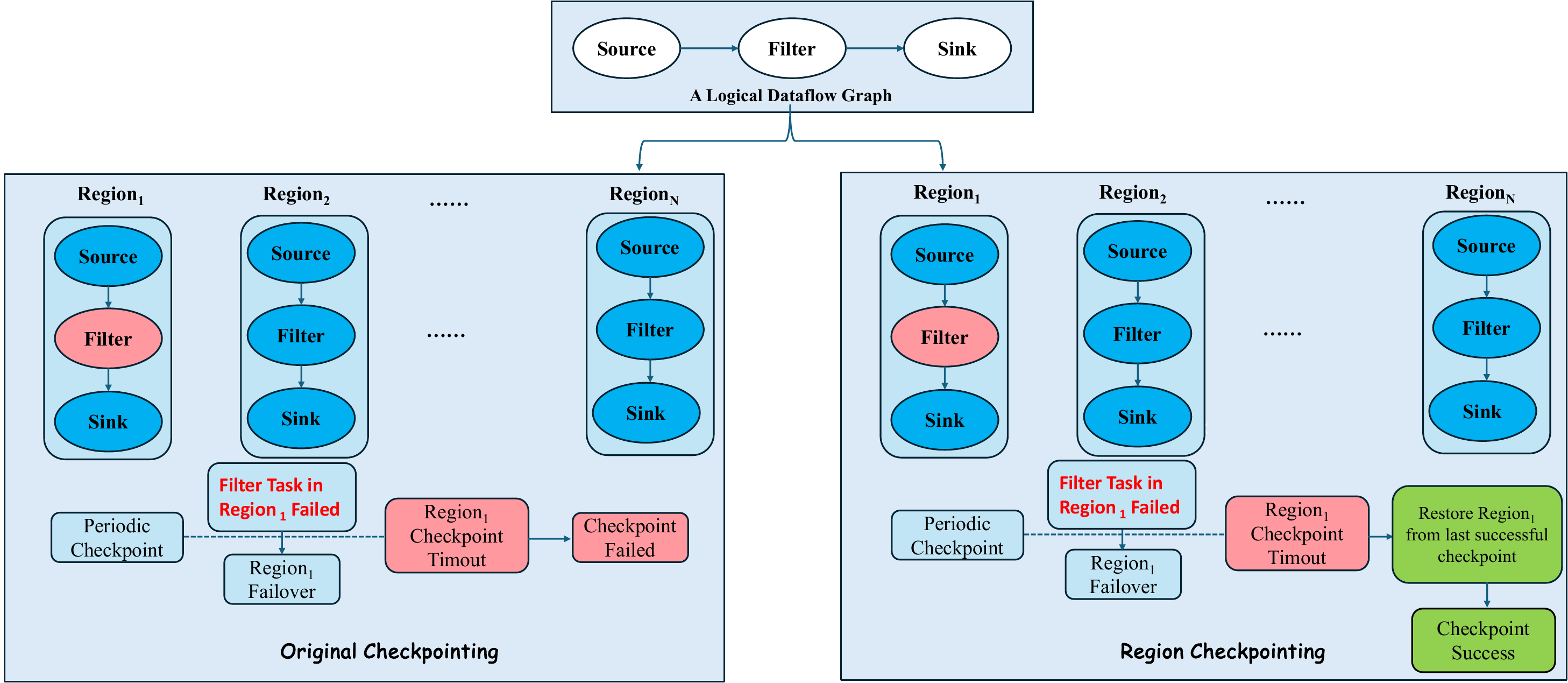}
  \caption{Original v.s. Region Checkpointing.}
 \vspace{-2em}
  \label{fig:regionCP}
\end{figure*}

\noindent \uline{\textbf{AutoScaling.}}
Large-scale stream processing workloads exhibit highly dynamic traffic patterns, often driven by user activity fluctuations, diurnal cycles, or bursty events. 
Over-provisioning resources based on peak demand is wasteful and leads to low utilization, whereas under-provisioning degrades service quality by increasing backlog and latency.
Traditional autoscaling approaches~\cite{mei2020turbine,floratou2017dhalion,castro2013integrating,xu2016stela} rely on coarse metrics such as CPU or throughput, which fail to capture the nuanced interplay between input rates, operator selectivity, and state size~\cite{kalavri2018three,fu2017drs,lohrmann2015elastic}. 
This mismatch often results in reactive scaling decisions that either lag behind workload shifts or oscillate between over- and under-provisioning.
\sysname adopts an autoscaling strategy inspired by the DS2 model~\cite{mao2023streamops,kalavri2018three}.
Instead of scaling based solely on resource utilization, DS2
identifies scaling bottlenecks by analyzing operator-level signals such as input rate, processing rate, backlog growth, and operator fan-out.

\revise{However, applying DS2 in production-grade environments requires addressing two key limitations of the original design: its reliance on stable metrics and its assumption of unconstrained reconfiguration.
First, to improve model accuracy under metric distortion caused by transient fluctuations or workload skew, \sysname incorporates smoothing and compensation strategies. 
Specifically, short-term metric oscillations are averaged, busy-time estimates at sources are adjusted with correction factors, and true processing rates are substituted for saturated busy-time signals under backpressure.
Second, to further improve stability, \sysname integrates rollback and safety mechanisms.
These include automatic rollback upon failed adjustments, business-driven shrinkage policies (e.g., disabling downscaling during peak hours), 
and protective measures such as rate limiting and failover-aware circuit breaking to avoid cascading effects.
Collectively, these refinements overcome DS2’s sensitivity to noisy metrics and instability under large-scale reconfiguration, enabling reliable operation in production-grade streaming deployments.}

While DS2 performs effectively in most scenarios, it often fails to deliver optimal operator parallelism when confronted with more complex streaming topologies of streaming jobs.
As a natural extension, we envision integrating StreamTune~\cite{streamtune2025icde}, a learning-based autoscaling framework that continuously refines scaling policies through historical workload traces and online feedback.
Unlike heuristic or rule-based approaches, StreamTune utilizes a pre-training and fine-tuning framework that leverages global knowledge from historical execution data for job-specific parallelism tuning.
Notably, it utilizes the Graph Neural Network~\cite{DBLP:conf/iclr/XuHLJ19,DBLP:conf/nips/HamiltonYL17,xu2018representation} to capture the complex correlation between the operator parallelism, DAG structure, and the identified operator-level bottlenecks.

%% file: sections/fault_tolerance.tex
\subsection{Fine-Grained Fault Tolerance}\label{sec:fault_tolerance}

Traditional fault-tolerance strategies in Apache Flink guarantee correctness but, due to their coarse granularity, often incur significant recovery latency and unnecessary disruption in large-scale stateful workloads.
To strengthen the resiliency of our production streaming cluster, we introduce a suite of fine-grained fault-tolerance mechanisms in \sysname that refine the recovery process.

\noindent \uline{\textbf{Region checkpointing.}}
\revise{In large-scale production environments, the original global checkpointing mechanism in Apache Flink often becomes a limiting factor for system resiliency. For instance, in a typical data synchronization workload, a single streaming job may consist of tens of thousands of tasks, each processing a distinct partition from an upstream message queue such as ByteMQ~\cite{mao2024bytemq}. Assuming checkpoints are triggered every five minutes and each task has a failure probability of 0.0001, it is
expected that at least one task will fail during every checkpoint interval. 
Under Flink’s original checkpointing scheme, the failure of any single task causes the ongoing checkpoint to be aborted.
This behavior leads to frequent checkpoint instability in large streaming jobs. The resulting inefficiency not only wastes computational resources but also undermines system resiliency, increasing recovery latency and elevating the risk of violating SLOs.}

To overcome this limitation, \sysname introduces region checkpointing, which isolates localized failures by partitioning checkpoints by execution regions\footnote{Here, a region represents a failure-recovery unit within the job execution graph, defined as a set of interconnected tasks bounded by blocking data exchanges such that tasks within the same region mutually depend on one another for recovery.}.
It is well-suited to large, heterogeneous streaming jobs that exhibit high operator parallelism and varied state sizes. 
Specifically, when the checkpoint of a specific region fails, the system automatically merges the latest successful checkpoint of that region with the current successful checkpoints from other regions, producing a consistent global checkpoint.
Figure~\ref{fig:regionCP} illustrates the difference in checkpoint behavior under task failures between the original and region checkpointing.
Starting from a logical dataflow graph with three operators, the physical execution graph is partitioned into $N$ disjoint regions.
When a task in Region$_1$ fails, both the original checkpointing scheme (left) and region checkpointing (right) trigger task failover and job recovery.
The key difference lies in how checkpoint coordination handles partial failures.
In the original scheme, because Region$_1$ fails to complete its task-level checkpoints, the entire checkpoint attempt is marked as failed.
In contrast, with region checkpointing, the system automatically merges the latest successful checkpoint state of Region$_1$ with those from other regions (e.g., Region$_2$ through Region$_N$), resulting in a globally successful checkpoint.
This mechanism effectively improves the overall checkpoint success rate by preventing a single regional failure from invalidating the entire checkpoint.

\noindent \uline{\textbf{Single-task Recovery.}}

To narrow the recovery scope, the Apache Flink community introduced the Region Failover strategy~\cite{flink_task_failure_recovery}, which restarts only the failed region and its dependent operators rather than the entire job.
Although the Region Failover strategy guarantees global state consistency, it can still incur substantial recovery latency.
\revise{In latency-sensitive scenarios where a small amount of data loss is acceptable, this coarse-grained approach may further introduce noticeable service jitter,
causing unnecessary disruption to otherwise unaffected operators.}

To address such scenarios, \sysname introduces a mechanism called \textit{Single-task Recovery}.
This mechanism addressed the above inefficiencies by localizing the recovery scope to the failed task, without cascading restarts to its enclosing region. 
Upon detecting a task failure, the JobManager initiates recovery only for that task, while all other tasks in the topology continue processing. This design effectively decouples fault handling from the global failover chain, thereby reducing both the restart scope and the state restoration workload. 
Note, this mechanism trades strict exactly-once guarantees for reduced downtime and lower operational impact, making it well-suited for latency-sensitive applications with relaxed consistency requirements.

Technically, the mechanism redefines the communication protocol between the failed task and its neighbors. 
For upstream operators, records destined for the failed task are dropped during its downtime; once the task resumes, normal data transmission is re-established. 
For downstream operators, we discard any incomplete or partial outputs produced before the failure, ensuring that subsequent processing resumes from a clean state. 
This selective data purging and channel reconnection allow the failed task to reintegrate into the job without triggering failover of all tasks within the same region. 
Besides, it minimizes interference with ongoing computation while maintaining predictable operator-level semantics.

Single-task Recovery is particularly effective \revise{for sample stitching workloads in large-scale recommendation pipelines, where data fragments are associated via key hashing to reconstruct complete training samples for the underlying machine learning models. As a small degree of sample loss has negligible impact on the recommendation model's accuracy, single-task recovery achieves a balanced trade-off between latency and correctness.}
Empirical evidence from production deployments shows that, compared to Region Failover, this mechanism reduces recovery latency by orders of magnitude, mitigates backlog accumulation, and stabilizes throughput under frequent localized failures. As a result, it offers a practical and resource-efficient pathway to sustaining SLOs in low-latency streaming environments.

\noindent \uline{\textbf{State Lazyload.}}
In Flink’s conventional recovery workflow, operator state is restored by downloading checkpoint data from remote storage and reinitializing the corresponding state backend (e.g., RocksDB~\cite{dong2021rocksdb}) before data processing can resume.
This eager state restoration model often results in substantial downtime when state sizes are large or when remote storage systems exhibit stragglers.

To address this limitation, \sysname introduces State LazyLoad, which decouples job resumption from full state materialization.
Instead of blocking recovery until all states are locally available, operators resume execution with partial states while the remaining portions are asynchronously restored in the background.
Technically, this mechanism transforms recovery into a demand-driven process. 
Upon task restart, operators promptly reestablish dataflow pipelines and resume processing with the subset of state already available locally.
When state entries not yet materialized are accessed, asynchronous retrieval is triggered from remote storage to ensure correctness without blocking execution.
By overlapping state restoration with normal processing, the mechanism minimizes downtime and reduces backlog growth.

State LazyLoad is particularly effective in large-scale stateful jobs with state sizes reaching tens of terabytes, as well as in heterogeneous storage environments where straggler nodes slow down recovery.
In such scenarios, this mechanism significantly reduces recovery-induced interruption, thereby improving resiliency for latency-sensitive workloads such as financial transactions, online recommendation pipelines, and real-time monitoring systems.

%% file: sections/startup.tex
\subsection{Job Startup Acceleration}\label{sec:startup}

\begin{table}[t]
  \caption{Job Startup Overhead of Nexmark Q2}
	\resizebox{\columnwidth}{!}{
		\begin{tabular}{c|c|c|c}
			\hline
            \rowcolor{mygrey}
			 \begin{tabular}[l]{@{}c@{}}  \textbf{\# of TM}  \\ \end{tabular} & \begin{tabular}[l]{@{}c@{}} \textbf{Job Parsing} \\ (ms) \end{tabular} & \begin{tabular}[l]{@{}c@{}} \textbf{Resource Allocation} \\ (ms) \end{tabular} & \begin{tabular}[l]{@{}c@{}} \textbf{Task Deployment} \\ (ms) \end{tabular} \\ \hline
			512 & 315 & 234,977 & 9,446 \\ \hline
            1024 & 268 & 338,615 & 15,565 \\ \hline
			2048 & 800 &\textbf{500,889} & \textbf{35,476}  \\ 
			\hline
	\end{tabular}}
	\vspace{-1em}
	\label{tab: job_startup_analysis}
\end{table}
To meet stringent latency and recovery time requirements, \sysname incorporates a set of mechanisms to accelerate job startup and restarts.
In production streaming environments, job restarts are often triggered by abnormal events, such as node failures, or planned maintenance, such as business logic updates and system upgrades.
Faster startup narrows the duration during which the job is unavailable or operating in a degraded state, reducing the risk of SLO violations. 

Table~\ref{tab: job_startup_analysis} reports the job startup overhead of Nexmark Q2~\cite{nexmark2024} under a fixed parallelism of two per TM, decomposed into job parsing, resource allocation, and task deployment across varying numbers of TMs. 
All three stages contribute noticeable latency, with job parsing taking several hundred milliseconds and resource allocation and task deployment incurring increasingly significant delays as the cluster grows.
At 2048 TaskManagers, the total startup latency grows to several hundred seconds.
These results highlight that job startup remains expensive, revealing significant optimization opportunities in large-scale environments.

\noindent \uline{\textbf{Optimizing Job Parsing and Deployment.}}
To reduce job startup overhead, \sysname applies optimizations in both the job parsing and task deployment stages.
During job parsing, a \textit{memory object reuse} mechanism is introduced in the JobManager to avoid redundant instantiation of execution plan objects, with focus on edge objects in the dataflow graph that represent data communication between tasks.
The mechanism identifies identical or semantically similar edges in the physical execution plan, such as those sharing the same partitioning scheme
or shuffle strategy, and reuses them instead of creating new instances.
In addition to reducing the total number of execution plan objects, this approach minimizes the memory footprint of individual objects by consolidating metadata and task description information, thereby lowering serialization costs in the subsequent deployment phase.

In the task deployment phase, traditional approaches typically dispatch computational tasks to TaskManagers (TMs) individually, which incurs substantial communication overhead and latency.
To mitigate this inefficiency, \sysname employs a \textit{batching strategy} in which 
the JobManager aggregates deployment requests for multiple tasks assigned to the same TM into a single consolidated message before transmission. By batching all task deployment descriptors destined for the same TM, this approach significantly reduces the number of remote procedure calls (RPCs) and network round-trips between the JobManager and TaskManagers, thereby lowering communication overhead and accelerating large-scale job deployment.

\noindent \uline{\textbf{Mitigating Slow-starting TaskManagers.}}
During job deployment, slow-starting TaskManagers can incur significant delays to job startup.
The JobManager must allocate and deploy tasks to all required TMs before transitioning the job into the next \textit{running} state.
\revise{If the job fails to obtain sufficient resources in time and the TaskManager allocation is delayed, the overall job startup latency will increase.}

To mitigate this issue, \sysname adopts a two-step strategy.
In the first step, it monitors indicators such as registration latency and slot reporting time to identify slow-starting TaskManagers, classifying those that take substantially longer than their peers as stragglers.
\revise{The second step is triggered only when TaskManager allocation exceeds a configurable threshold (e.g., $2$ minutes),} upon which the JobManager proactively provisions up to 30\% additional TaskManagers, bounded by a configurable maximum (e.g., $5$), to ensure sufficient resources without waiting for all slow TaskManagers to complete initialization.
Once the job reaches the running state, these redundant TMs are promptly released to reduce resource overhead.
\revise{This optimization is particularly effective in \textit{containerized} environments where task initialization is slowed down by underlying node performance bottlenecks, such as I/O saturation on a physical machine that delays container image downloads and consequently prolongs job startup time.}

\noindent \uline{\textbf{HotUpdate.}}
To further speed up the job restarts, \sysname introduced a HotUpdate mechanism, designed for scenarios with frequent business logic updates.
In conventional restarts, the JobManager cancels the old job, releases TaskManager slots, and reacquires resources before redeployment, which can incur considerable startup latency.
HotUpdate eliminates this overhead by reusing allocated resources rather than entirely tearing down and reinitializing the execution environment.
Specifically, the old job is stopped directly on the existing TaskManagers, and the new job is deployed to the same running processes, thereby preserving slot allocations and JVM warm-up states.
Empirical evidence from our production environment demonstrates that, when combined with other startup acceleration techniques, HotUpdate can reduce the job restart latency to  20 seconds. 

From the resiliency perspective, the HotUpdate mechanism strengthens the system’s ability to withstand operational disruptions, particularly during frequent business logic updates or system maintenance. 
By eliminating the downtime associated with complete redeployment, it 
reduces the exposure window for potential data loss or backlog accumulation. Furthermore, 
It allows for rolling upgrades of business logic with minimal disturbance to upstream and downstream systems, effectively bridging the gap between high availability and rapid iteration in large-scale streaming environments. This makes HotUpdate a valuable tool for sustaining service stability in production-grade Flink clusters.

%% file: sections/high-availability.tex
\section{High-availability Construction}
\label{sec:high-availability}
Achieving high availability (HA) in large-scale Flink clusters requires fault-tolerance mechanisms that span both the execution layer and its external dependencies.
On one hand, replication strategies determine how quickly jobs can recover from task and operator failures, balancing recovery latency against resource efficiency across heterogeneous workloads. On the other hand, the reliability of critical external systems such as HDFS, Zookeeper, \godel directly impacts Flink’s ability to preserve exactly-once semantics and sustain continuous service. This section presents the high-availability construction of \sysname, which integrates hybrid replication with dependency-aware fault tolerance to ensure resilient and uninterrupted operation in production environments.

\subsection{Replication Strategy}

Replication strategies are central to achieving high availability and high resiliency in DSPSs.
There are two typical strategies called active replication and passive replication.
Active replication~\cite{balazinska2008fault,shah2004highly} maintains multiple concurrent replicas of an operator, each processing the same input stream in parallel.
This strategy enables rapid failover with negligible recovery latency, as standby replicas can immediately assume execution when a failure occurs. The primary drawback of active replication lies in its substantial resource overhead, along with the requirement that operators exhibit deterministic behavior to guarantee consistency across replicas.

In contrast to active replication, passive replication~\cite{gu2009empirical,hwang2006cooperative,kwon2008fault} achieves HA with lower resource overhead by maintaining standby replicas that remain idle until a failure occurs. Instead of executing in parallel with the primary, these replicas periodically receive state updates, which are applied during recovery. While this approach conserves computational resources, it incurs longer recovery latency since a failed operator must be restarted and its state reloaded before resuming execution. Despite this drawback, passive replication remains a practical strategy for workloads where strict low-latency recovery is not mandatory but resource efficiency is critical.

Shared with similar ideas to prior work~\cite{zhang2010hybrid,heinze2015adaptive,su2016tolerating}, \sysname adopts a hybrid approach to balance recovery latency and resource efficiency across heterogeneous business scenarios. 
Building on Flink’s native support for periodic checkpoint,
passive replication serves as our default mechanism, offering fault tolerance with minimal resource overhead for lower-priority workloads such as dashboard monitoring and analytical pipelines.
For latency-sensitive applications with strict availability requirements,
\sysname selectively switches to active replication through pre-deployed secondary copies to ensure rapid failover and uninterrupted service continuity. 
By aligning replication modes with workload priorities, this hybrid strategy achieves a practical balance between efficiency and responsiveness, thereby strengthening the overall resiliency of large-scale stream processing systems.

\subsection{Fault Tolerance with External Dependencies}

As a stateful stream processing framework, our internal Flink cluster relies on several external systems to provide fault tolerance and high availability.
Core dependencies include HDFS for durable state management~\cite{borthakur2007hadoop,malhotra2025evaluating}, Apache Zookeeper for cluster coordination and JobManager leader election~\cite{hunt2010zookeeper,li2021rt}, and G{\"o}del for resource orchestration and task deployment~\cite{xiang2023godel,wang2025capsys}. 
These components are tightly integrated into Flink’s execution pipeline, such that their failures can directly disrupt job scheduling, or impair cluster availability. 
Unlike transient task failures, which can often be mitigated through localized recovery, failures in these dependency systems manifest as systemic disruptions that threaten the reliability of all running jobs.
The complexity of diagnosing and handling such failures is amplified in large-scale production environments, where operational dependencies span multiple layers of storage, coordination, and scheduling. 
This subsection analyzes the failure modes of these systems in our production deployment and presents the fault-tolerance mechanisms developed to mitigate their impact.

\noindent \uline{\textbf{HDFS Fault Tolerance}}: 
HDFS serves as an important persistent storage for Flink checkpoints and savepoints, making its reliability fundamental to system resiliency.
Typical failure modes include NameNode crashes and DataNode unavailability. 
Such failures can directly cause checkpoint timeouts, trigger alignment delays that induce backpressure, and result in repeated job restarts.

To mitigate these risks, \sysname mandates that HDFS be deployed in a highly available configuration with active–standby NameNodes and replication across multiple DataNodes to avoid single points of failure.
To guard against long-tail storage latencies and localized failures, \sysname combines incremental
checkpoints with region-level recovery and lazy state restoration, thereby reducing recovery scope and I/O pressure during failover as mentioned in Section~\ref{sec:fault_tolerance}.
In addition, augmenting HDFS with alternative durable storage backends (e.g., object storage) provides resilience against prolonged HDFS outages.
Finally, continuous health monitoring and automated failover procedures are incorporated to detect anomalies promptly and to trigger rapid recovery actions, thereby strengthening overall system resiliency and robustness.

\noindent \uline{\textbf{Zookeeper Fault Tolerance}}: 
Apache Zookeeper
serves as a core coordination service for Flink, supporting cluster metadata management and leader election.
Its stability directly affects JobManager's HA and task scheduling.
However, it is susceptible to failures such as session expirations, quorum loss, leader election storms, and high-latency operations under server overload.
These issues may lead to frequent JobManager leadership changes, stalled job submissions, or misdiagnosed task failures, threatening cluster resiliency. 

\revise{To reduce the strong dependency on ZooKeeper, \sysname introduces a lightweight fault-tolerance mechanism that prevents running jobs from failing when ZooKeeper becomes unavailable.
Specifically, \sysname maintains a redundant copy of the leader metadata in HDFS in addition to ZooKeeper.
Upon detecting a ZooKeeper failure, the system automatically falls back to the HDFS copy to retrieve the leader information, ensuring uninterrupted job execution.
Only when both ZooKeeper and HDFS are unavailable, or when the leader information stored in HDFS becomes inconsistent with the in-memory state, will the affected jobs be terminated to preserve correctness.}

  \begin{figure*}[tb]
  \centering
  \includegraphics[width=0.75\linewidth]{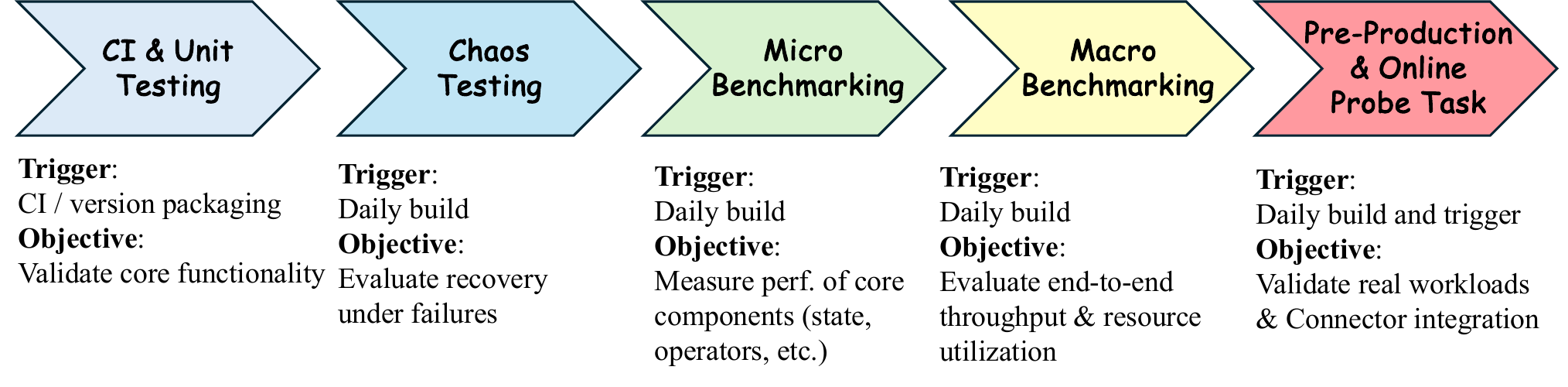}
  \caption{Testing and Release Pipeline of the Flink Engine.}
 \vspace{-1em}
  \label{fig:release}
\end{figure*}

 \noindent \uline{\textbf{G{\"o}del Fault Tolerance}}: 
\godel functions as the resource orchestration and task deployment framework for Flink, responsible for allocating computing resources, launching TaskManagers, and dynamically scaling operator parallelism at runtime.
Its reliability is essential for ensuring timely job initialization and stable task execution. 
Typical failure modes include scheduler unavailability or throttling, and stale or inconsistent resource views that cause resource misallocation.
Such failures can manifest as delayed task deployments, prolonged recovery times, or slow-starting TMs that increase startup latency.

To mitigate these risks, \sysname incorporates multiple layers of fault tolerance. 
At the engine level, \sysname integrates optimizations such as HotUpdate that enable restarting jobs without releasing existing resources, and the slow-starting TM mitigation technique, where up to 30\% redundant TMs are allocated to absorb startup delays and promptly released once the job is running (see Section~\ref{sec:startup}). 
At the control-plane level, redundant orchestration endpoints and fallback strategies are deployed to mask transient \godel failures~\cite{mao2023streamops}. 
Deployment requests incorporate exponential backoff~\cite{kwak2005performance} and idempotent retries~\cite{ramalingam2013fault} to avoid duplicated submissions under unstable conditions.
\revise{In particular, when job submission fails due to temporary \godel unavailability, \sysname automatically retries with exponential backoff and performs \textit{job uniqueness validation} to prevent duplicate executions arising from repeated submissions.}
Specifically, exponential backoff prevents excessive load during transient failures by spacing out retries, while idempotent retries guarantee correctness by ensuring repeated requests do not cause duplicate effects.
In addition, health checks and continuous monitoring are employed to detect degraded scheduler states and trigger automated remediation. Collectively, these mechanisms ensure that orchestration-layer instabilities do not escalate into systemic disruptions, thereby preserving the resiliency of large-scale Flink deployments.

\section{Robust Testing and Release Pipeline}\label{sec:release}
From an engineering standpoint, improving the resiliency of a stream processing engine also requires systematic validation across its critical components.
To this end, \sysname employs a robust testing and release pipeline that continuously exposes the engine to fault-prone and high-load conditions, enabling early detection of latent defects and ensuring stable releases.
By uncovering vulnerabilities prior to production deployment, this pipeline strengthens the system’s ability to withstand runtime failures.
This section presents the pipeline overview, followed by detailed discussions of two representative testing methods: chaos testing and performance testing.

\subsection{Overview}

Figure~\ref{fig:release} illustrates \sysname's testing and release pipeline of the Flink engine, which integrates multi-layer validation to ensure both functional correctness and production reliability. 
The process is initiated with \textit{continuous integration (CI) and unit testing}, triggered upon version packaging, where automated test suites verify core functionalities and ensure backward correctness. 
Building upon this foundation, \textit{chaos testing} is executed on daily builds to examine the system’s resilience under diverse failure scenarios.

To complement fault-resilience validation, \textit{micro benchmarking} is conducted regularly to measure the performance of core components such as operator execution and state backends. These fine-grained benchmarks provide insights into the efficiency of critical components and highlight potential performance regressions introduced by code changes. In parallel, \textit{macro benchmarking} offers macro-level evaluation by deploying synthetic streaming queries at cluster scale, enabling the measurement of end-to-end throughput and resource utilization under realistic workloads.

Finally, pre-production and online probe tasks are continuously executed on daily builds to validate the engine’s stability under real-world business scenarios and its interoperability with external dependencies. Probe tasks are lightweight streaming jobs that emulate representative workloads and interact with external services, 
thereby exercising end-to-end data paths without incurring significant overhead. This stage bridges controlled experimental testing with live production environments, ensuring that each release meets both system-level performance objectives and business-level resiliency requirements prior to large-scale deployment.

\begin{table*}[ht]
  \caption{Workload Selection.}
  \vspace{-0.5em}
	\resizebox{2\columnwidth}{!}{
		\begin{tabular}{l|l|l|l}
			\hline
            \rowcolor{mygrey}
			\textbf{Job} & \textbf{Description} & \textbf{Logical Dataflow Graph} & \textbf{Source Rate}  \\ \hline
		Nexmark Q2 & \begin{tabular}[l]{@{}l@{}}Filters bid records that satisfy  predefined conditions. \end{tabular} & 
        \begin{tabular}[l]{@{}l@{}} Two Logical Nodes\\ with one Source \end{tabular}  & Fixed at 0.8 M/s \\ \hline
			Nexmark Q12 & 
            \begin{tabular}[l]{@{}l@{}} Count bids per user within a fixed processing-time window \end{tabular}
            &  \begin{tabular}[l]{@{}l@{}} Three Logical Nodes\\ with one Source \end{tabular} & Fixed at 0.8 M/s \\ \hline
			Data Synchronization (DS)& 
            \begin{tabular}[l]{@{}l@{}} Ingests records from A Message Queue into the Apache Hive \end{tabular}
            & \begin{tabular}[l]{@{}l@{}} Two Logical Nodes\\ with one Source \end{tabular} & \begin{tabular}[l]{@{}l@{}} Varies between\\ 1 M/s-6.4 M/s \end{tabular} \\ \hline
            Sample Stitching (SS) & 
            \begin{tabular}[l]{@{}l@{}} Real-time stitching of samples based on the online \\ recommendation outcomes\end{tabular}
            & \begin{tabular}[l]{@{}l@{}} Dozens of Logical Nodes\\ with two Sources\end{tabular} & \begin{tabular}[l]{@{}l@{}}Varies between \\ 5 K/s-45 K/s \end{tabular}
            
            \\
			\hline
	\end{tabular}
    }
	\vspace{-2em}
	\label{tab: exp_workload}
\end{table*}

\subsection{Chaos Testing}
Chaos testing is a systematic fault-injection methodology in distributed systems, where controlled failures are deliberately introduced to evaluate a system’s resilience to unexpected disruptions and its capacity to maintain service continuity under adverse conditions.
In \sysname, chaos testing is employed as a critical validation tool for Flink’s fault-tolerance mechanisms.
At the hardware level, we simulate anomalies such as increased network latency, bandwidth throttling, CPU interference, and disk I/O contention to capture the effects of resource heterogeneity and transient infrastructure bottlenecks. 
At the process level, we proactively terminate TaskManager and JobManager instances during active job execution to emulate real-world failure scenarios such as node crashes or container evictions.
These tests systematically examine the system’s ability to recover execution state from checkpoints, reassign resources, and restore operator pipelines without violating correctness guarantees. By combining hardware-level perturbations with process-level failures, chaos testing provides comprehensive coverage of both transient and catastrophic disruptions, offering insights into recovery latency, checkpoint reliability, and the preservation of end-to-end job semantics in large-scale DSPS deployments.

\subsection{Performance Testing}
To ensure system robustness and guard against performance regressions during Flink upgrades,  \sysname establishes a structured performance testing framework that covers both component-level and system-level evaluation.
This framework is divided into two complementary categories: micro benchmarking and macro benchmarking, each serving distinct diagnostic and validation purposes.

\noindent \uline{\textbf{Micro Benchmarking}}: 
Micro benchmarking focuses on the critical computational paths of the Flink engine.
Specifically, we evaluate three major aspects: \textit{operator execution}, \textit{job scheduling}, and \textit{state management}.
Notably, for state management, we investigate the latency of read/write operations, scalability under large key-space expansion (i.e., scenarios where the number of distinct keys maintained in operator state grows substantially), and recovery efficiency across diverse checkpointing configurations.
All evaluations are conducted on single-node environments to minimize external variability and ensure reproducibility.
The primary goal is to identify performance regressions in individual components before they propagate to end-to-end workloads.

\noindent \uline{\textbf{Macro Benchmarking}}: Macro benchmarking provides an end-to-end assessment of Flink’s performance under workload scenarios that approximate production-grade streaming applications.
In \sysname, we employ the Nexmark suite~\cite{nexmark2024}, which comprises a collection of continuous queries modeled on an auction system. 
These queries span diverse processing patterns, including source ingestion, filtering, aggregation, and more complex windowed and join-based operations.
Benchmarks are executed on a multi-node Flink cluster, where system-level metrics such as CPU utilization and query throughput are continuously monitored. 
The collected results are analyzed through interactive dashboards to capture workload dynamics and guide both performance tuning and release validation.

%% file: sections/experiment.tex
\begin{figure*}[tbp]
    \centering
    \vspace{-0.5em}
    \begin{subfigure}[b]{0.32\textwidth}
        \centering
        \includegraphics[width=\textwidth]{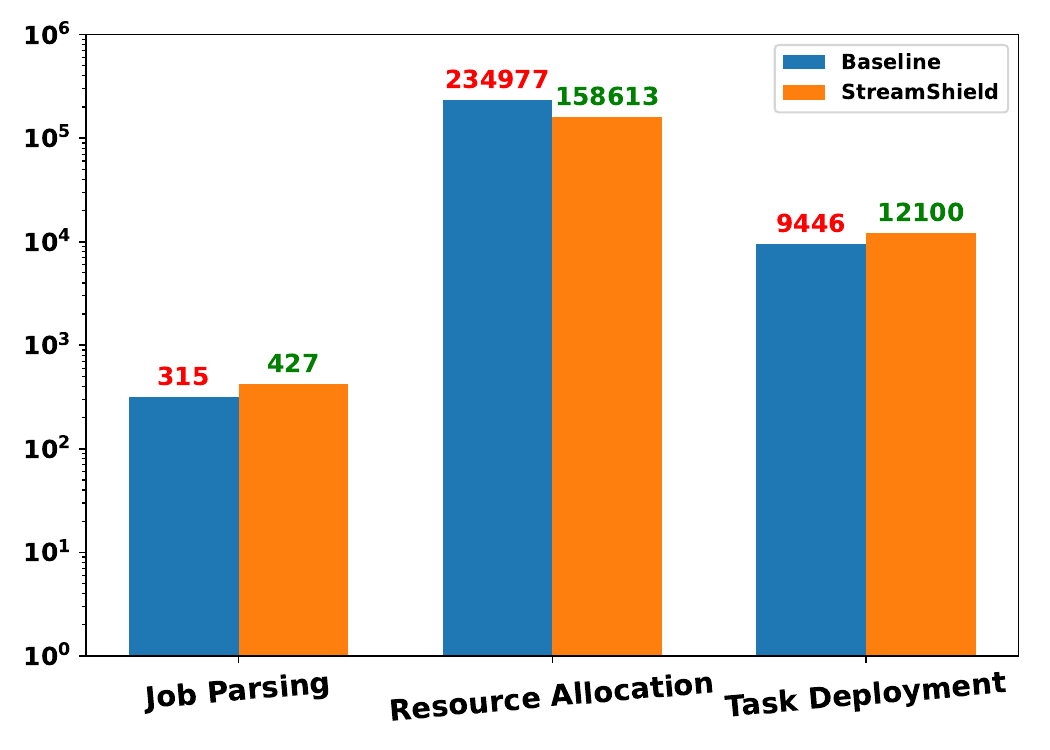}
        \caption{512 TMs}
        \label{fig:256TMs}
    \end{subfigure}
    \hfill
    \begin{subfigure}[b]{0.32\textwidth}
        \centering
        \includegraphics[width=\textwidth]{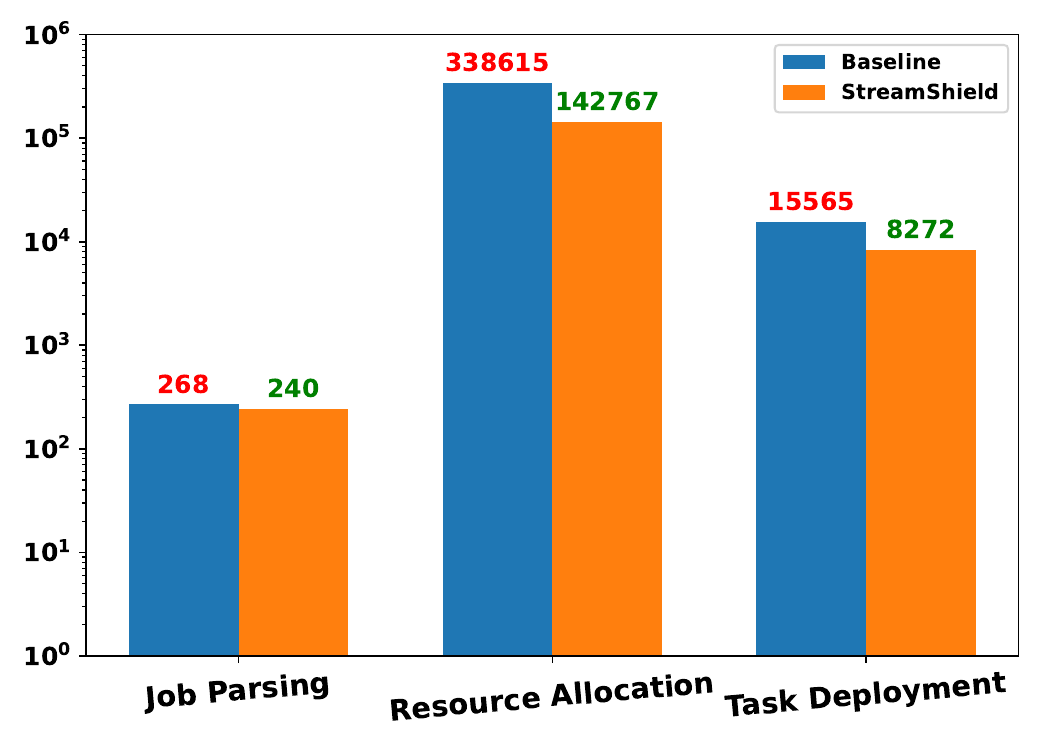}
        \caption{1024 TMs}
        \label{fig:512TMs}
    \end{subfigure}
    \hfill
    \begin{subfigure}[b]{0.32\textwidth}
        \centering
        \includegraphics[width=\textwidth]{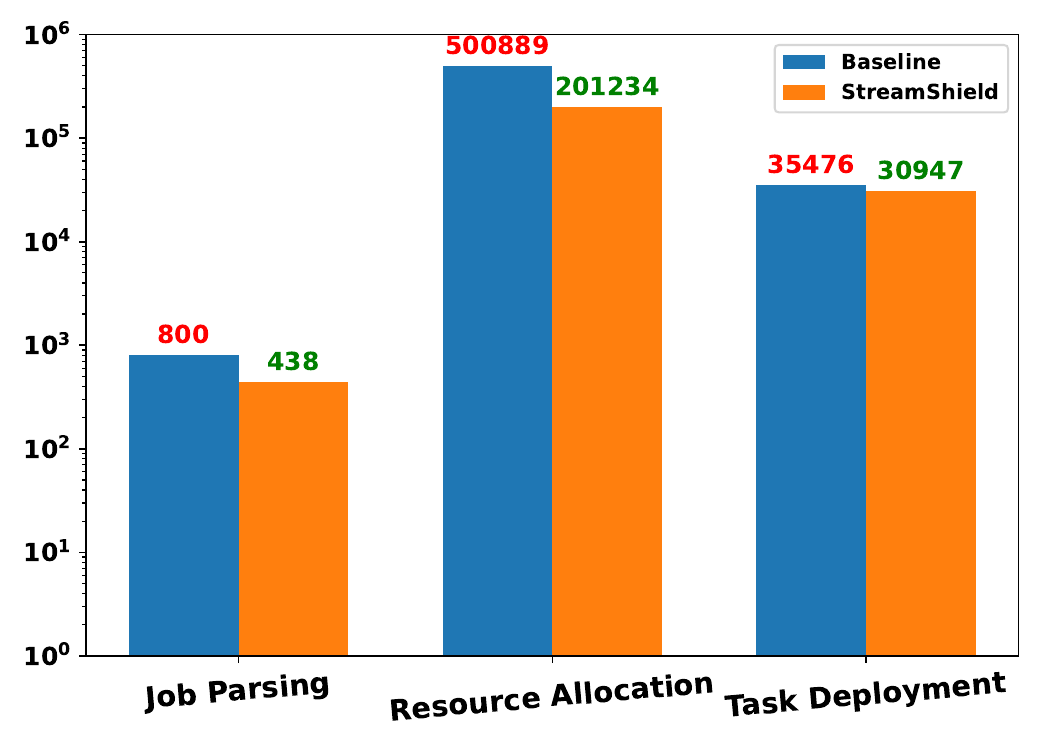}
        \caption{2048 TMs}
        \label{fig:2048TMs}
    \end{subfigure}
    \vspace{-0.5em}
    \caption{Comparison of Job Startup Overhead Across Different Numbers of TMs.}
     \vspace{-1.5em}
    \label{fig:exp_startup}
\end{figure*}

\section{Experimental Evaluation} \label{sec:experiment}
This section presents the performance evaluation of \sysname conducted in ByteDance’s production environment.
We focus on representative engine-oriented mechanisms 
to demonstrate the efficiency and effectiveness of \sysname.
We first introduce the experimental setup, and then examine the impact of 
each
proposed technique.

\subsection{Experimental Setup}
\noindent \textbf{Environment.}

We deploy \sysname on a \godel-managed Flink v1.11 cluster deployed at ByteDance.
The cluster provides 384 AMD Genoa CPU cores, 2.3 TB of DRAM, and 15 TB of NVMe SSD storage, interconnected through dual 100 Gbps network interfaces.
Flink is configured with 16 GB process memory for both the JobManager and TaskManager, where each TaskManager is allocated four CPU cores and four task slots.
The system employs the RocksDB state backend with incremental checkpointing,
and performs checkpointing every 60 seconds.

\noindent \textbf{Workload.}
Table~\ref{tab: exp_workload} summarizes the representative workloads used in our evaluation.
Here, K/s and M/s denote thousands and millions of records per second, respectively.
Nexmark Q2 contains a simple stateless filter and Q12 introduces stateful sliding-window aggregation; both operate on compact logical DAGs with fixed source rates of 0.8 M/s.
Data Synchronization (DS) reflects a production-critical ingestion pipeline that transfers records from a message queue into Apache Hive, modeled as a two-node DAG with input rates varying from 1 M/s to 6.4 M/s. 
Sample Stitching (SS) represents a more complex workload, characterized by dozens of logical nodes and dual input streams, where the source rate fluctuates between 5 K/s and 45 K/s.

\subsection{Job Startup Overhead Comparison}

We evaluate the effectiveness of \sysname in accelerating job startup using Nexmark Q12, a representative complex query. 
The experiment scales the number of TaskManagers (TMs) while fixing two task slots per TM.
We compare the baseline system with \sysname in terms of job startup overhead, breaking down the results into three distinct phases: job parsing, resource allocation, and task deployment.
In this evaluation, we replace SSD storage in the experimental cluster with HDD-based storage to emulate realistic online performance bottlenecks.

Fig.~\ref{fig:exp_startup} compares the job startup overhead of the baseline and \sysname across different cluster scales. At 512 TMs (Fig. 5a), both systems incur minimal parsing overhead (315 ms vs. 427 ms), but resource allocation dominates startup cost, with the baseline requiring 234,977 ms compared to 158,613 ms in \sysname. Task deployment shows a similar trend, where \sysname slightly outperforms the baseline (12,100 ms vs. 9,446 ms).
At 1024 TMs (Fig. 5b), parsing costs remain small (268 ms vs. 240 ms), while the baseline’s resource allocation overhead rises to 338,615 ms, more than twice that of \sysname (142,767 ms). Deployment overhead also diverges, with the baseline reaching 15,565 ms compared to 8,272 ms under \sysname.
At the largest scale of 2048 TMs (Fig. 5c), the baseline suffers from severe resource allocation delays (500,889 ms), whereas \sysname reduces this cost to 201,234 ms. Parsing overhead is modest in both cases (800 ms vs. 438 ms), while task deployment overhead is again lower in \sysname (30,947 ms vs. 35,476 ms).

Overall, the results demonstrate that resource allocation is the dominant contributor to job startup overhead, and \sysname substantially mitigates this bottleneck. As the cluster size grows, \sysname reduces resource allocation overhead by more than 50\% and improves task deployment efficiency, leading to significantly faster job initialization at scale.

\begin{figure}[tb]
  \centering
  \vspace{-1em}
  \includegraphics[width=0.75\linewidth]{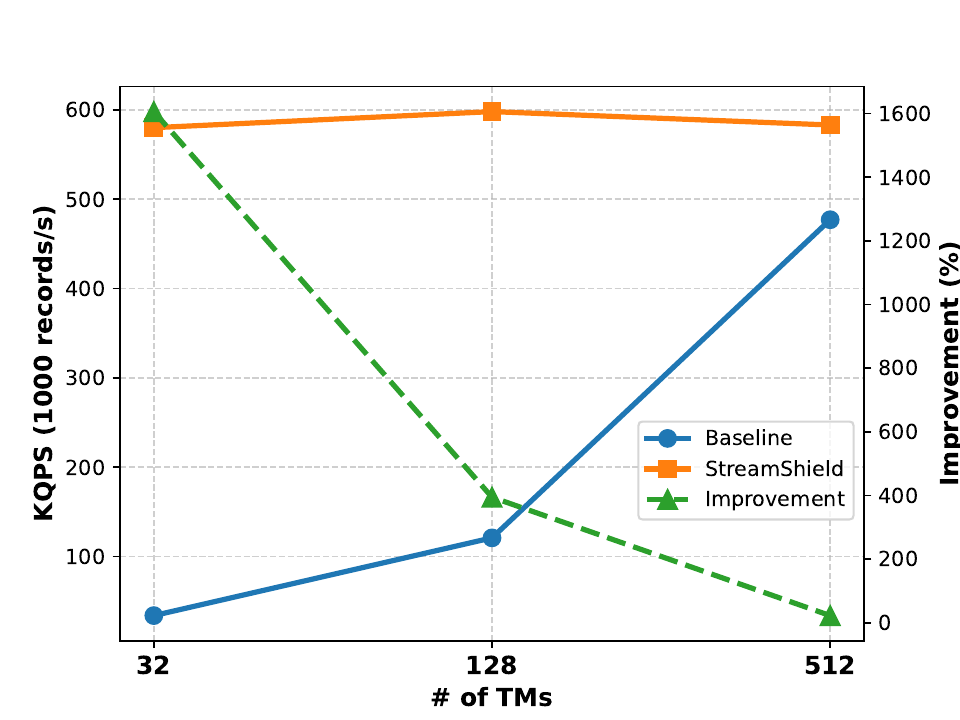}
  \vspace{-1em}
  \caption{QPS w/ and w/o Adaptive Shuffle.}
 \vspace{-1.5em}
  \label{exp:adaptive_shuffle}
\end{figure}

\subsection{Effect of Adaptive Shuffle}
To evaluate the effectiveness of Adaptive Shuffle, we select Nexmark Q2, which involves two logical operators.
To introduce workload imbalance, 10\% of the tasks in the second operator are randomly delayed by 1000 ms before processing each input record.
The operator parallelism is fixed at 8, and we compare the throughput of the baseline with that of \sysname enabled with the backlog-based shuffle mechanism proposed in Section~\ref{sec: adaptive_shuffle}.
As shown in Fig.~\ref{exp:adaptive_shuffle}, \sysname consistently outperforms the baseline across cluster scales. 
At 32 TMs, the baseline suffers from severe skew, achieving only 34 KQPS, while \sysname sustains 580 KQPS, yielding more than 1600\% improvement.
At 128 TMs, the baseline reaches 121 KQPS compared to 598 KQPS with \sysname, an improvement of approximately 400\%.
At 512 TMs, the baseline reaches 477 KQPS, narrowing the performance gap, but \sysname still achieves 583 KQPS.

The relatively steady QPS achieved by \sysname with backlog-based shuffle can be attributed to its ability to dynamically balance workload distribution at runtime. 
In the baseline, \revise{the static round-robin shuffle strategy} leads to skewed data assignment, where a subset of tasks receives disproportionately more records and quickly becomes a bottleneck. 
This imbalance forces the entire pipeline to slow down, resulting in sharp throughput degradation under load skew.
By contrast, backlog-based shuffle continuously monitors the downstream operators’ backlog status and dynamically redistributes records away from congested tasks toward less-loaded ones, effectively alleviating stragglers and reducing queuing delays.
As a result, \sysname avoids abrupt performance collapses observed in the baseline and sustains stable throughput across varying cluster sizes, demonstrating the robustness of adaptive shuffle against skew-induced workload imbalance.

\begin{figure}[tb]
  \centering
  \includegraphics[width=0.75\linewidth]{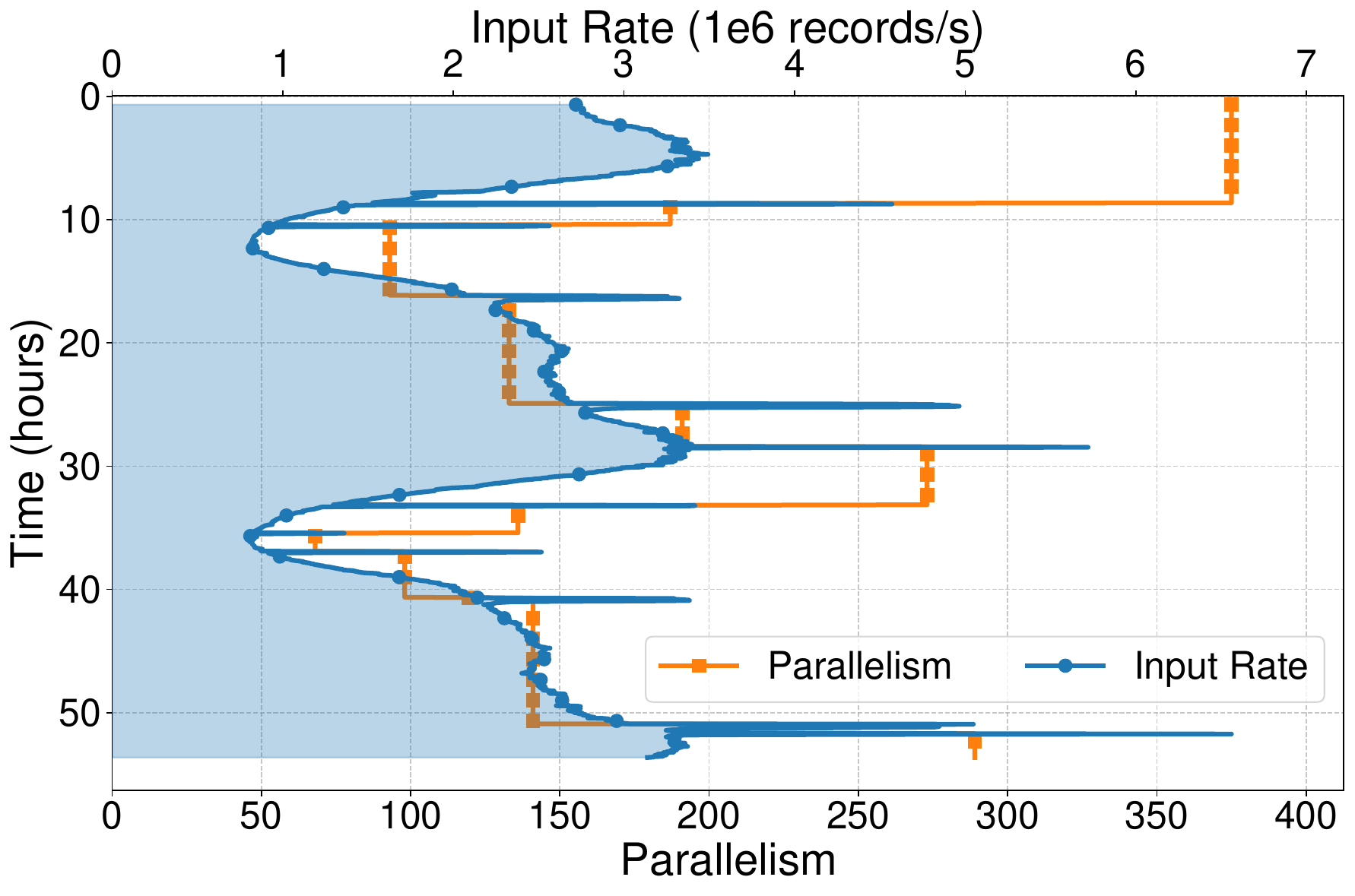}
  \vspace{-0.5em}
  \caption{AutoScaling under Variable Input Rates.}
    \vspace{-1em}
  \label{exp: auto_scaling}
\end{figure}

\subsection{Effect of Autoscaling}
To evaluate the effectiveness of the proposed autoscaling technique in \sysname, we select the Data Synchronization (DS) job as a representative workload.
During execution, the input rate is varied continuously, and we observe how \sysname adapts by adjusting the parallelism of a logical operator within the job.
Fig.~\ref{exp: auto_scaling} illustrates the dynamic behavior of autoscaling in \sysname under fluctuating workloads. Over the 55-hour execution period, the input rate (blue curve, top axis) varies significantly, oscillating between 1 and nearly 7 million records per second. At the beginning, the input rate grows steadily, and \sysname reacts by increasing the parallelism from around 150 to 200, ensuring sufficient processing capacity. Around the 15-hour mark, the input rate drops sharply, and the parallelism is reduced accordingly, thereby avoiding unnecessary resource occupation. A similar pattern occurs between 20 and 35 hours, where bursts in input rate trigger rapid scale-up of parallelism to nearly 200, followed by scale-down when the workload recedes. Between 40 and 55 hours, the system faces another surge, with input rate approaching 7M/s; in response, parallelism is raised to nearly 300, demonstrating the system’s ability to elastically provision large-scale resources. After the peak, parallelism is gradually decreased as the input rate subsides, highlighting the mechanism’s efficiency in releasing resources once the demand diminishes.

Overall, the figure shows that \sysname's autoscaling technique maintains a strong correlation between input rate and allocated parallelism. The system responds promptly to workload bursts by scaling up parallelism to preserve throughput, while scaling down when the input rate declines to conserve resources. This adaptive adjustment not only mitigates performance bottlenecks under heavy load but also prevents resource over-provisioning during light load, thereby achieving both efficiency and elasticity in long-running streaming jobs.

\begin{figure}[tb]
  \centering
  \includegraphics[width=0.75\linewidth]{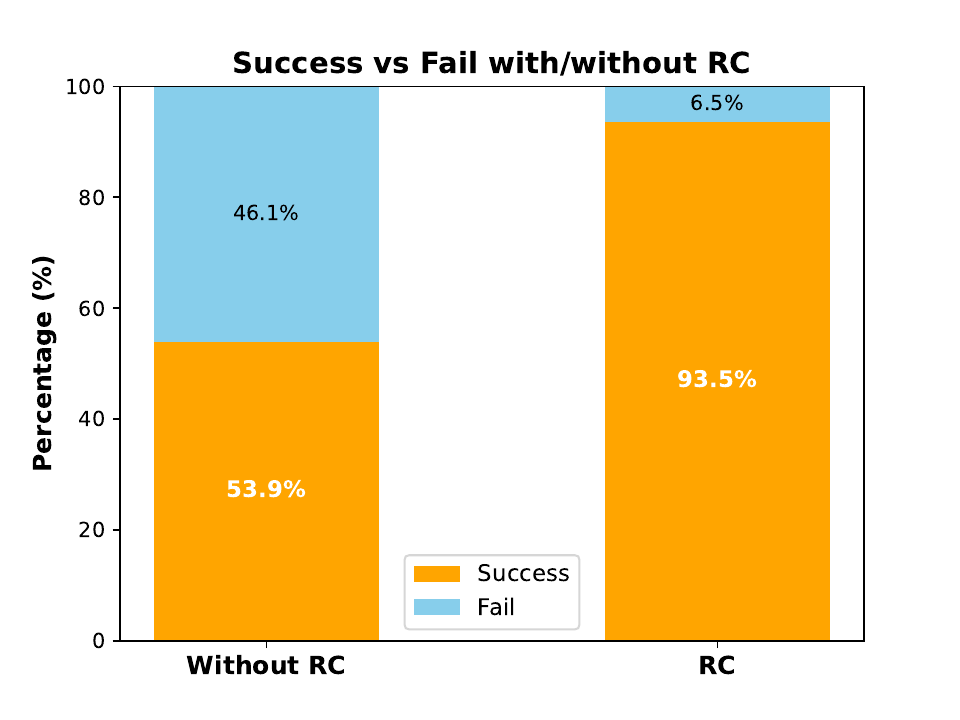}
  \vspace{-1.6em}
  \caption{Checkpoint Success Rate w/ and w/o RC.}
 \vspace{-2em}
  \label{exp:region_checkpointing}
\end{figure}

\subsection{Effect of Region Checkpointing}
To evaluate the effectiveness of Region Checkpointing (RC), we employ the DS job, a representative workload that accounts for nearly 10\% of our internal streaming applications.
It is critical to determine whether RC can provide tangible benefits for this representative class of jobs.
To emulate realistic instability, we randomly inject a 5\% probability of slow HDFS uploads during task execution, while configuring the checkpoint interval to 30 seconds. Each experiment runs for 12 hours, ensuring that the system undergoes a sufficient number of recovery events to allow for a controlled and reproducible comparison between configurations with and without RC.
This setup ensures a controlled and reproducible evaluation environment and allows us to directly assess the benefits of RC in improving Flink’s checkpoint reliability in distributed settings.

Fig.~\ref{exp:region_checkpointing} reports the checkpoint success rate with and without RC under the DS workload. Without RC, only 53.9\% of checkpoints complete successfully (1253 out of 2326), while 46.1\% fail due to simulated slow HDFS uploads. In contrast, enabling RC significantly improves reliability, with 93.5\% of checkpoints succeeding (2099 out of 2244) and only 6.5\% failing. This stark difference highlights the robustness of RC in mitigating the cascading effects of partial failures. By isolating failure domains and retrying at the region level, RC prevents global checkpoint aborts and sustains a much higher completion rate. These results confirm that RC is highly effective in improving checkpoint reliability for large-scale data synchronization jobs, which are sensitive to frequent state persistence.

\begin{figure}[tb]
  \centering
  \includegraphics[width=0.75\linewidth]{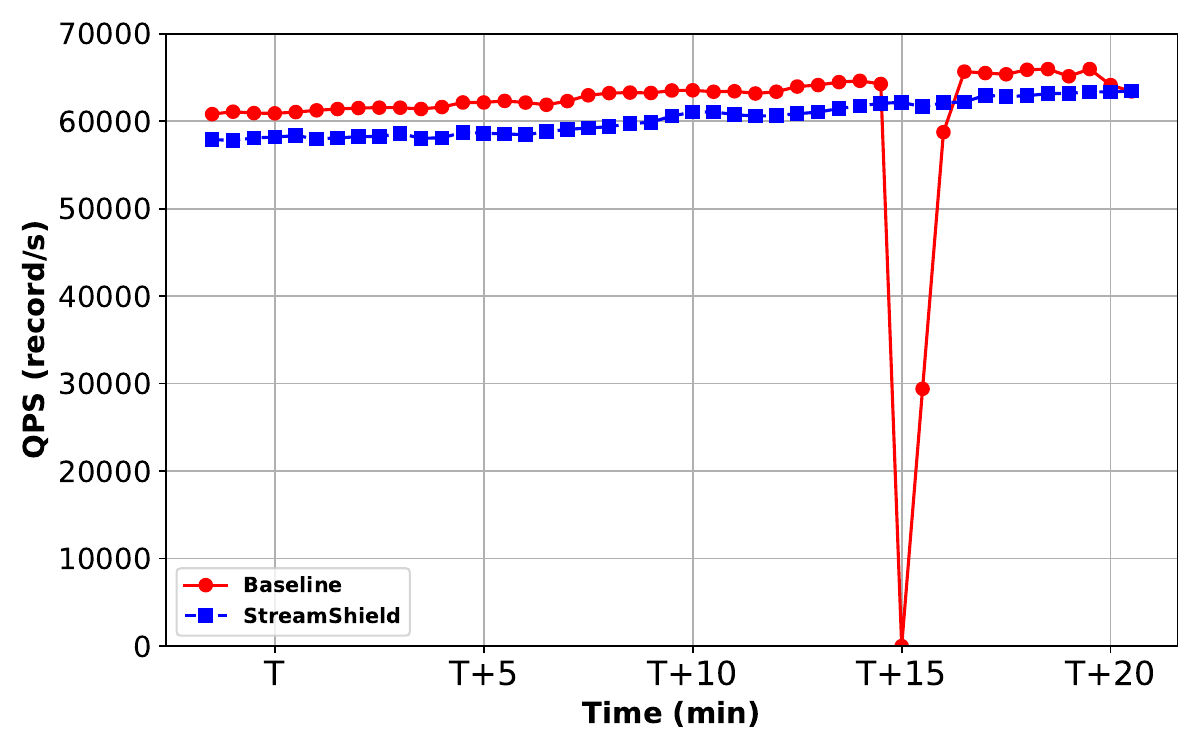}
  \vspace{-1em}
  \caption{QPS under TaskManager Failure.}
 \vspace{-2em}
  \label{exp:single_task}
\end{figure}

\subsection{Effect of Single-task Recovery}
For the evaluation of single-task recovery, we select the SS job, which requires real-time result delivery while tolerating only minor data loss. 
The job performs a dual-stream join in a video recommender system, combining feature data used for generating recommendations with post-recommendation behavioral feedback (e.g., click events or watch duration).

To assess the impact of failures, we deliberately kill one TaskManager during execution and monitor the job’s throughput (QPS) under two settings: the baseline and \sysname with single-task recovery, as shown in Fig.~\ref{exp:single_task}.
In the baseline, the failure injected at approximately T+15 minutes forces the entire job to restart, which
leads to a complete drop in throughput to zero, followed by a gradual recovery process that lasts several minutes, during which the service remains unavailable.
\revise{This is because the stitching workload involves a \textsf{Join} operator with an all-to-all data exchange pattern between upstream and downstream tasks, where the native failover mechanism propagates a single task failure to all connected tasks, triggering region-wide recovery.}
Such behavior is detrimental in real-time systems, where even short interruptions can significantly degrade user experience.

In contrast, \sysname enables single-task recovery and isolates the failed subtask without triggering a global restart. As a result, the job sustains nearly steady throughput throughout the failure event, with only marginal and short-lived fluctuations around T+15. This fine-grained recovery mechanism eliminates the need to reinitialize unaffected tasks, thereby substantially reducing recovery time and mitigating performance disruption. Overall, the experiment demonstrates that single-task recovery provides strong fault tolerance for real-time streaming workloads, ensuring continuous service availability even in the presence of TaskManager failures.

%% file: sections/conclusion.tex
\section{Conclusion}

This paper presents \sysname, a resiliency framework built on one of the world’s largest Apache Flink clusters at ByteDance, which is carefully designed along engine, cluster, and release perspectives.
\sysname integrates fine-grained fault tolerance, runtime optimization, hybrid replication, and dependency-aware mechanisms to address failures at both the execution layer and in external systems.
In addition, a rigorous testing and release pipeline validates these mechanisms under fault-prone and high-load scenarios before deployment, further strengthening production resiliency.

Future work will focus on proactive diagnosis and recovery, leveraging machine learning to adjust operator parallelism and detect abnormal events before they affect business-critical applications.